\documentclass[aps,twocolumn,prc,epsfig]{revtex4}
\usepackage{graphicx}
\usepackage{epsfig}
\def\be{\begin{eqnarray} &&}

\def\ee{\end{eqnarray}}
\def\psla{\rlap \slash} 
\input psfig.sty
\begin{document}
\title{Weak decay constant of pseudoscalar mesons in a 
QCD-inspired model\footnote{{\bf To appear "Braz. J. of Phys.
(2003).}}}
\author{L. A. M. Salcedo$^{a}$, J.P.B.C. de Melo $^{b}$, 
D. Hadjmichef$^{c}$, T.
 Frederico$^{a}$}
\address{
$^a$Dep. de F\'\i sica, Instituto Tecnol\'ogico de Aeron\'autica,
Cento T\'ecnico Aeroespacial, 12.228-900 S\~ao Jos\'e dos Campos,
S\~ao Paulo, Brazil. \\
$^{b}$Instituto de F\'{i}sica Te\'{o}rica, Universidade Estadual
Paulista, 01405-900, S\~{a}o Paulo, Brazil.
\\
$^{c}$ Instituto de F\'\i sica e Matem\'atica, Universidade
Federal de Pelotas, 96010-900, Campus Universit\'ario Pelotas, Rio
Grande do Sul, Brazil.}

\date{\today}
\begin{abstract}
We show that a linear scaling between the  weak decay constants of
pseudoscalars and the vector meson masses is supported by the
available experimental data. The decay constants scale as
$f_m/f_\pi=M_V/M_\rho$ ($f_m$ decay constant and $M_V$ vector
meson ground state mass). This simple form  is justified within a
renormalized light-front QCD-inspired model for quark-antiquark
bound states.
\end{abstract}
\maketitle

\section{Introduction}

Effective theories used to describe hadrons, which are inspired by
Quantum Chromodynamics\cite{pauli0,pauli2,pauli3} can be useful in
indicating direct correlations between observables of different
hadrons. Therefore, it is possible to pin down the relevant
dependence of the observables with some physical scales  that
otherwise would have no simple reason to present a direct
relation, besides being properties of the same underlying theory.
For example, a systematic dependence of a hadron observable with
its mass can offer useful guide for presenting results obtained in
Lattice QCD. Systematic correlations between different meson
properties with mass scales are also found from the solution of
Dyson-Schwinger equations\cite{cr}.

One intriguing aspect is the dependence of the weak decay constant
of the pseudoscalar meson with its mass. For light mesons up to
$D$, the weak decay constant tends to increase with the mass,
while numerical simulations of quenched lattice-QCD indicate that
$f_D > f_B$\cite{flynn}, which is still maintained with two flavor
sea quarks \cite{flynn,c}. General arguments, within
Dyson-Schwinger formalism for QCD in the heavy quark limit, says
that the weak decay constant should be inversely proportional to
$\sqrt{M_m}$\cite{cr1} ($M_m$ is the pseudoscalar mass). Effective
QCD inspired models valid for low energy scales can also be called
to help to investigate this subtle point. In these models
\cite{pauli2,pauli3,tobpauli}, the interaction is flavor
independent, while the masses of constituent quarks can be
changed, which naturally implies in  correlations between
observables and masses.

Our aim here, is to investigate the pseudoscalar weak decay
constant within a QCD inspired model \cite{tobpauli}. The
effective mass operator equation for the lowest Light-Front
Fock-state component of a bound system of a constituent quark and
antiquark of masses $m_1$ and $m_2$, obtained in the effective
one-gluon-exchange interaction approximation \cite{pauli0} and
simplified in the $\uparrow\downarrow$-model\cite{pauli2,pauli3}
to
\begin{eqnarray}
M_m^2\psi_m(x,{\vec k_\perp})= \left[\frac{{\vec
k_\perp}^2+m^2_1}{x}+\frac{{\vec k_\perp}^2+m^2_2}{1-x}
\right]\psi_m (x,{\vec k_\perp})
\nonumber \\
-\int dx' d{\vec k'_\perp} \xi(x,x') 
\left(\frac{4m_1m_2}{3\pi^2}\frac{\alpha}{Q^2}-\lambda\right)
\psi_m (x',{\vec k'_\perp}), \label{p1}
\end{eqnarray}
where the phase space factor is
$$\xi(x,x')= \frac{\theta(x')\theta (1-x')} {\sqrt{x(1-x)x'(1-x')}}~,$$
and $\psi_m$ is the projection of the light-front wave-function in
the quark-antiquark Fock-state. The mean square momentum transfer
 $((k^\prime_1-k_1)^2+(k^\prime_2-k_2)^2)/2$ gives $-Q^2$ ($k_i$ and
$k^\prime_i$ are the quark four-momenta). The coupling constant
$\alpha$ defines the strength of the Coulomb-like potential and
$\lambda$ is the bare coupling constant of the Dirac-delta
hyperfine interaction. The energy transfer in $Q^2$ is left out.
Confinement comes through the binding of the constituents in the
meson, which in practice keeps the quarks inside the mesons.

The mass operator equation (\ref{p1}) needs to be regularized and
renormalized in order to give physical results, such development
has been performed in Ref.\cite{tobpauli}. In that work, it was
obtained the renormalized form of the equation for the bound state
mass, which is  $i$) invariant under renormalization group
transformations, $ii$) the physical input is given by the pion
mass and radius $iii$) no regularization parameter.

In the work of Ref.\cite{tobpauli}, the quark mass  was changed to
allow the study of mesons with one light antiquark plus a strange,
charm or bottom quark. The masses of the constituent quarks were
within the range of 300 up to 5000 MeV. For the up-down quarks a
mass of 384 MeV is found from the rho meson mass, which in the
model is weakly bound.  The Dirac-delta interaction comes from an
effective hyperfine interaction which splits  the pseudo-scalar
and vector meson states. In the singlet channel the hyperfine
interaction is attractive, which is not valid for the spin one
mesons. In the model, the Dirac-delta interaction mock up
short-range physics which are brought by the empirical value of
the pion mass, and a reasonable description of the binding
energies of the constituent quarks forming the pseudoscalar mesons
is found \cite{tobpauli}. The model, without the Coulomb like
interaction, was also able to describe the binding energies of the
ground state of spin 1/2 baryons containing two light quarks and a
heavy one\cite{suisso}.

Within the effective model of Eq.(\ref{p1}), the low-lying vector
mesons are weakly bound systems of constituent quarks while the
pseudo-scalars are more strongly bound \cite{tobpauli}. This
allows to calculate the masses of the constituent quarks  directly
from the masses of the vector mesons ground states\cite{suisso}:
\begin{eqnarray}
 m_u &=&\frac{1}{2}M_{\rho}=384{\ MeV}  \ ,\nonumber \\
 m_s&=&M_{K^{*}} - \frac{1}{2}M_{\rho}=508{\ MeV} \ , \nonumber \\
 m_c&=&M_{D^{*}} - \frac{1}{2}M_{\rho}= 1623{\ MeV} \ , \nonumber \\
 m_b &=&M_{B^{*}} - \frac{1}{2}M_{\rho}=4941{\ MeV}\ ,
\label{mconst}
\end{eqnarray}
where it is used the values of 768 MeV, 892 MeV, 2007 MeV and 5325
MeV for the $\rho$, $K^*$, $D^*$ and $B^*$ masses,
respectively\cite{pdg}.

Here, we use the effective model to predict a physical property
directly related to the wave-function of the ground state of the
pseudo scalar mesons. We calculate the weak decay constants
($f_m$) of $K^+$, $D^+$, $D^+_s$, for which experimental values
are known \cite{pdg}. Besides the constituent quark masses from
Eq.(\ref{mconst}) and the pion mass, our calculation needs as
input the pion weak decay constant, $f_\pi=\ 92.4 \pm 0.07\pm
0.25$ MeV\cite{pdg}. The eigenfunction of the interacting mass
squared operator from Eq.(\ref{p1}) for large transverse momentum
behaves as the asymptotic wave-function, which decreases slowly as
$ p^{-2}_\perp$. Therefore, in the calculation of the weak decay
constants it is necessary to regulate the logarithmic divergence
in the transverse momentum integration and take care of the
cut-off dependence to be able to give an unique answer. One has to
consider that the pion decay constant provides the short-range
information contained in the pion wave function, which we suppose
to be the same for all pseudo-scalars. Here, we just write the
divergent integral in the transverse momentum in terms of $f_\pi$
and from that obtain the other decay constants.

\section{Meson light-front wave function}

 The wave function of the meson ($m$) is the solution of
Eq.(\ref{p1}). In the approximation where the Coulomb-like
interaction is considered in lowest order, the pseudo-scalar meson
wave function is given by\cite{tobpauli}
\begin{eqnarray}
&&\psi_m(x,\vec k_\perp)= {1\over \sqrt{x(1-x)}}{G_m\over
M_m^2-M_0^2} \nonumber \\ && \times \left[ 1 -   \int \frac{dx'
d{\vec k'_\perp}\theta(x')\theta (1-x')}{\sqrt{x'(1-x')}}
\left(\frac{4m_1m_2}{3\pi^2}\frac{\alpha}{Q^2}\right)\right.
\nonumber \\ && \left. \times {1\over M_m^2-{M'}_0^2} \right] ,
\label{piphi}
\end{eqnarray}
where
\begin{eqnarray}
M_0^2=\frac{{\vec k_\perp}^2+m^2_1}{x}+\frac{{\vec
k_\perp}^2+m^2_2}{1-x}, \label{m0}
\end{eqnarray}
in the frame in which the meson has zero transverse momentum.
(${M'}_0^2$ is obtained from $M^2_0$ by substitution of $\vec
k_\perp$ and $x$ by $\vec k'_\perp$ and $x'$, respectively.) The
overall normalization of the $q\overline q$ Fock-component of the
meson wave-function (\ref{piphi}) is $G_m$.

In this first calculation of the decay constant within this model,
we are going to assume the dominance of the asymptotic form of the
meson wave function and simply use
\begin{eqnarray}
\psi_m(x,\vec k_\perp)={1\over\sqrt{x(1-x)}}{G_m\over M_m^2-M_0^2}
\ . \label{mwf}
\end{eqnarray}

To obtain the pseudoscalar decay constants, we follow
Ref.\cite{mill}. To construct the observables in terms of the
meson wave function, one has to account for the coupling of the
quark spins, which is described by an effective Lagrangian density
with a pseudo-scalar coupling between the quark ($q_1(\vec x)$ and
$q_2(\vec x)$) and meson $\left(\Phi_m(\vec x)\right)$
fields\cite{mill}
\begin{eqnarray}
{\cal L}_{eff}(\vec x)= - i G_m \Phi_m(\vec x) \  \overline
q_1(\vec x) \gamma^5 q_2(\vec x) + h.c.\ , \label{lag}
\end{eqnarray}
the coupling constant is $G_m$. From the effective Lagrangian
above one can derive meson observables and write them in terms of
the light-front asymptotic wave function, Eq. (\ref{mwf}). To
achieve this goal, it is necessary to eliminate the relative
$x^+$-time ($x^+=t+z$) between the constituents in the physical
amplitude, which then allows  to write the meson observable in
terms of the wave function \cite{mill}.

\section{Results for the weak decay constant of pseudoscalar mesons}

The pseudoscalar meson weak decay constant is calculated from the
matrix element of the axial current $A^\mu (0)$, between the
vacuum state  $|0\rangle$ and the meson state $|q_m\rangle $
 with four  momentum $q_m$ \cite{pdg}:
\begin{eqnarray} \
\langle 0 \mid A^\mu( 0)\mid q_m\rangle= \imath \sqrt{2} f_m
q_m^\mu  \ , \label{fm}
\end{eqnarray}
where $A^\mu(\vec x)=  \overline q(\vec x) \gamma^\mu \gamma^5
q(\vec x)$.

Using the pseudoscalar Lagrangian, Eq. (\ref{lag}), one can
calculate the matrix element of the axial current, which is
expressed by a one-loop diagram and can be written as:
\begin{eqnarray} &&\imath \sqrt{2} M_m f_m= \nonumber \\&& N_c G_m
\int \frac{d^4k}{(2\pi)^4} Tr\left[ \gamma^+\gamma^5
S_2(k)\gamma^5 S_1(k-q_m) \right] , \label{fm1}
\end{eqnarray}
where $\gamma^+=\gamma^0+\gamma^3$, $N_c=3$ is number of colors
and $S_i(p)=\imath/(\psla p -m_i+\imath \epsilon)$ is the
propagator of the quark field.

\begin{center}
\begin{table}[t,b,h]
\caption { Results for the pseudoscalar meson weak decay constants
$f_m$ calculated with Eq.(\ref{fim}). The inputs for the model are
$m_q$, $m_{\overline q}$, $M_m$ and $f_\pi$ given in the table.
All masses and decay constants are in MeV. ( $^\dagger$ The
experimental mass of $D^{*+}_s$ is quite near to the model $c
\overline s$ vector meson mass which is given by $m_c+m_s=$ 2131
MeV. )}
\begin{tabular}{|c|c|c|c|c|}
\hline $q \overline q$ & $M_{m}$\cite{pdg} &$M_{v}$\cite{pdg}
&$f^{model}_{m}$ &$f_{m}^{exp}$\cite{pdg}
\\ \hline
$\pi^+ (u\overline d)$& 140 & 771 ($\rho$)& 92.4 & $92.4\pm.07\pm 0.25$   \\
$K^+(u \overline s) $& 494 & 892 ($K^*$)& 107 & $113.0\pm1.0\pm0.31$   \\
$ D^+(c \overline d)$& 1869& 2010 ($D^{*+}$)& 241 &$212^{+127+6}_{-106-28}$\\
$D^+_s(c \overline s)$& 1969& 2112$^\dagger$ ($D^{*+}_s$)& 253 & $201\pm13\pm28$\\
\hline
\end{tabular}
\end{table}
\end{center}

By integration over $k^-$ in Eq.(\ref{fm1}), the relative
light-front time between the quarks is eliminated and one obtain
the expression of $f_{m}$ suitable for the introduction of the
meson light front wave-function. So, performing the Dirac algebra
and integrating analytically  over $k^-$, one obtains
\begin{eqnarray}
f_m = -\frac{\sqrt{2}}{8\pi ^{3}}N_{c}\int^1_0 dx \left(
(1-x)m_{2}+xm_{1}\right) \nonumber \\ \times \int dk_{\perp
}^{2}{G_m\over{x(1-x)M^2_m-k^2_\perp-m_1x-m_2(1-x)}} ~,
\label{fm2}
\end{eqnarray}
in the meson rest-frame. We have used the momentum fraction
$x=k^+/q^+_m$.

One can write Eq.(\ref{fm2}) in terms of the valence component of
the pseudoscalar meson wave function as:
\begin{eqnarray}
f_m = &&\frac{\sqrt{2}}{8\pi
^{3}}N_{c}\int^1_0{dx\over\sqrt{x(1-x)}}\left(
(1-x)m_{2}+xm_{1}\right) \nonumber \\ && \times \int dk_{\perp
}^{2}\psi_m(x,\vec k_\perp) \ . \label{fm3}
\end{eqnarray}
The above expression is general and one can use it to  calculate
the decay constant of any  pseudoscalar meson state, and as well
one can use it to normalize the eigenfunction of the squared mass
operator from the solution of Eq.(\ref{p1}).

We observe that Eq.(\ref{fm2}), written in terms of the asymptotic
part of the valence wave function has a logarithmic divergence in
the transverse momentum integration due to the slow decrease of
the wave function. From a physical point of view, one could think
that the regularization scale is larger than the masses of the
quarks and the divergent transverse momentum integration, will be
defined through  the value of $f_\pi$, for example. Therefore, one
has:
\begin{eqnarray}
f_m = {\text {const.}}\int^1_0 dx \left( (1-x)m_{2}+xm_{1}\right)
\ , \label{fm4}
\end{eqnarray}
and $const.$ determined by $f_\pi$. One observe as well that, $f_m
\propto m_1+m_2$, which in our model is the vector meson mass,
thus one immediately gets:
\begin{eqnarray}
{f_m\over f_\pi} =  {M_v\over M_\rho}  \ . \label{fim}
\end{eqnarray}

The numerical results of Eq.(\ref{fim}) are shown in Table I. It
is verified that a reasonable description of the weak decay
constants of the pseudoscalar mesons is possible within the
effective light-front model. However, we have made use only of the
asymptotic form of the wave function and one needs to investigate
the decay constant with more refined wave functions, eigenstates
of the squared mass operator, Eq.(\ref{p1}), which includes the
dynamics of the effective quarks. Therefore, the results which are
overestimating  the heavier meson decay constants, can be an
indication that a more elaborated wave function is needed,
although one cannot discard that   other mechanisms could  be
relevant\cite{cr1}.

Also, we intend to perform the evaluation of the weak decay
constants using a more sophisticated version of the model, where
confinement is included \cite{conf},   which so far was shown to
describe the meson spectrum.

In summary, we have shown the existence of a direct
proportionality between the weak decay constants and the masses of
 the vector mesons ground states, which can be an useful tool in
 the systematic study of these quantities.

{\bf Acknowledgments:} We thank CNPq and FAPESP for financial
support.

\end{document}